\documentstyle[12pt]{article}
\input{epsf.tex}

\begin{document}

  \begin{flushright} \begin{small}
  DF/IST-8.2001 \\gr-qc/0207009
  \end{small} \end{flushright}


\begin{center}

{\bf The radial infall of a highly relativistic point particle into
a Kerr black hole along the symmetry axis}\\
\vskip 0.2cm
Vitor Cardoso \\
\vskip 0.1cm
{\scriptsize  CENTRA, Departamento de F\'{\i}sica,
	      Instituto Superior T\'ecnico,}\\ 
{\scriptsize  Av. Rovisco Pais 1, 1096 Lisboa, Portugal,}\\
{\scriptsize E-mail: vcardoso@fisica.ist.utl.pt}\\
\vskip 0.2cm

Jos\'e P. S. Lemos \\
\vskip 0.1cm
{\scriptsize  CENTRA, Departamento de F\'{\i}sica,
	      Instituto Superior T\'ecnico,} \\
{\scriptsize  Av. Rovisco Pais 1, 1096 Lisboa, Portugal,}\\
{\scriptsize E-mail: lemos@kelvin.ist.utl.pt}
\end{center} 

\begin{abstract}
In this Letter we consider the radial infall along the symmetry axis
of an ultra-relativistic point particle into a rotating Kerr black hole.
We use the Sasaki-Nakamura formalism to compute the waveform, energy
spectra and total energy radiated during this process. We discuss possible 
connections between these results and the black hole-black hole collision 
at the speed of light process.
\end{abstract}


\noindent
\section{Introduction}
\vskip 3mm

The lack of exact radiative solutions to Einstein's equations has
promoted perturbation theory in General Relativity into a special 
place, as the tool for analyzing gravitational
radiation emitted in physically interesting events.  The use of
perturbation theory in spacetimes containing black holes
started with the work of Regge and Wheeler \cite{regge}, where they
addressed the stability of the Schwarzschild geometry against small
deviations. This analysis was extended to include the infall of 
particles in a  Schwarzschild black hole 
by Zerilli \cite{zerilli} and others (see, e.g., 
\cite{davis}).

The Kerr geometry, without spherical symmetry, proved to be more
difficult to handle, but has also given some insights into relations
that previously seemed a mystery \cite{chandra}. Teukolsky
\cite{teukolsky} was able to decouple and separate the
perturbation equations for the Newman-Penrose quantities, and reduced
them to a single radial equation, now known as the Teukolsky equation.
However, it suffers from a couple of drawbacks, in that it has a long
range potential and a source term which is in general divergent at
large distances.  This makes it difficult to analyze radiation emitted
by test particles in generic orbits, though special cases have been
handled (see for example the case of circular orbits
\cite{detweiler}). Sasaki and Nakamura \cite{nakamurasasaki}
improved this situation by introducing a set of transformations that
bring the Teukolsky equation into the Sasaki-Nakamura  equation,
which has both a short range potential and an everywhere well behaved
source term. Moreover, their equation reduces to the Regge-Wheeler
equation when the rotation parameter is set equal to zero.  Using
their formalism, Sasaki, Nakamura and co-workers \cite{nmothers} have
computed the gravitational radiation for generic orbits of particles
falling, from rest at infinity, into a Kerr black hole. These studies
have recently been revisited \cite{hughes1,shibata} not only to study
the effects of radiation reaction, but also to produce accurate
waveforms, to serve as templates to the various gravitational wave
detectors already at work.  An extension of the Sasaki-Nakamura formalism to
perturbations other than gravitational is given in \cite{hughes2}.

It is important to stress that this approach is only justified as long
as $\mu$, the mass of the infalling particle, is much smaller than
$M$, the mass of the black hole, so that it can serve as perturbation
parameter.  By the late 70's however, when the first full numerical
simulations of black hole collisions were being done \cite{smarr}, it
became clear that taking the limit $\mu \rightarrow M$ gave
unexpectedly good results in perturbation theory.  
For recent improvements see \cite{anninos}.

The high velocity collision of two black holes, a problem interesting
in itself, has gained renewed interest recently, with the possibility
of black hole formation at TeV scales \cite{bhprod}. This process 
was studied extensively some time ago by D'Eath and Payne
\cite{payne}, by performing a perturbation expansion around the
Aichelburg-Sexl metric, which describes a Schwarzschild black hole
moving at the speed of light.  The question arises as to whether it is
possible to treat this process in the old Regge-Wheeler-Zerilli
approach. Is it possible to study the collision of two black holes
moving at the speed of light, by considering some expansion around a
Schwarzschild metric?  In a previous paper \cite{lemosvitor} we have
argued that it is possible, provided the agreement between
perturbation theory and numerical relativity is more than a
coincidence.  In spite of this unproven universal agreement, we do
have the strong conviction that it will hold in this process, but we
are ultimately justified by the excellent agreement between our
results \cite{lemosvitor} and previous studies \cite{payne}.  The
essence of our previous study was simple: consider a highly
relativistic particle falling into a Schwarzschild black hole, use the
Zerilli approach to compute the waveform and energy, Lorentz-boost the
black hole to high velocities in the direction of the infalling
particle, and end up with a collision at high velocity between a black
hole and a small test particle. If one employs the assumption that
this still works if $\mu \rightarrow M$ the result follows. Our
results \cite{lemosvitor} showed an excellent agreement with results
by D'Eath and Payne \cite{payne} and with results by Smarr
\cite{smarr2}, which leads us to believe that once again perturbation
theory gives very good results throughout all the values of the
perturbation parameter.  

One would now like to extend these results to the Kerr family of
rotating black holes, a less studied geometry.  D'Eath and Payne's
results do not apply here for example, although with probably a great
effort one could use their methods to study the collision of
Aichelburg-Sexl-Kerr particles, which solution has been found in
\cite{neugebaueretal}. It is true that we do not expect to see black
holes colliding at the speed of light in any astrophysical scenario,
but we do expect to see this at Planck energies, or, perhaps at the
LHC, should the TeV scenario \cite{bhprod} be correct.  In any case,
the holes will most probably be rotating.  Furthermore, the work done
so far in the Kerr geometry dealt only with infall which starts from
rest, so it will be interesting to see the outcome when the particle
has a non zero velocity at infinity, and compare it with results in
the Schwarzschild geometry.

Thus, the situation we consider in this paper is the following: a highly
relativistic point particle impinges radially into a Kerr black hole,
along its symmetry axis. We will use the Sasaki-Nakamura formalism to
find the energy radiated, which will be the main
result of this paper.  We will then Lorentz-boost the Kerr
hole to high velocity in the direction of the infalling object, which
basically amounts to put $M \rightarrow \gamma M $, where $\gamma$ is
the Lorentz factor.  This will then describe the high velocity
collision between a Kerr hole and a small particle.  
Finally, we put $\mu \rightarrow M $,
which means that we are dealing with the collision at nearly the speed of
light of two equal mass black holes, one rotating, the other
non-rotating.  The justification for doing this last step (not allowed
on formal basis) comes from the excellent results obtained so far by
perturbation theory \cite{davis,smarr,anninos,lemosvitor}.

\section{A Green's function solution to the Sasaki-Nakamura equation}
After some manipulations, the Teukolsky equation \cite{teukolsky} may
be brought to the Sasaki-Nakamura \cite{nakamurasasaki} form (details
about the Teukolsky formalism may be found in the original literature
\cite{teukolsky}, and also in \cite{breuerbook}. For a good account of
the Sasaki-Nakamura formalism we refer the reader to
\cite{nakamurasasaki},\cite{nmothers}, and \cite{hughes1}):
\begin{equation}
\frac{d^2}{dr_*^2} {X(\omega,r)}- {\cal F}\frac{d}{dr_*}{X(\omega,r)} - 
{\cal U} X(\omega,r) = {\cal L} \,.
\label{sn}
\end{equation}
The tortoise $r_*$ coordinate is defined by
$dr_*/dr=(r^2+a^2)/\Delta$, and ranges from $-\infty$ at the horizon
to $+\infty$ at spatial infinity.
The functions ${\cal F}$ and ${\cal U}$ can be found in the original
literature \cite{nakamurasasaki,nmothers}.
We are considering the radial infall of a highly relativistic particle
into a Kerr black hole along the symmetry axis, so the situation is
axisymmetric. This means that the azimuthal quantum number $m$
appearing in $\cal F$ and $\cal U$ \cite{nmothers}, may be set to
zero.  Also, in this simple situation it is possible to find an exact
form for the function ${\cal L}$ (compare this with the source term in
Schwarzschild \cite{lemosvitor}): 
\begin{equation}
{\cal L}=-\frac{\mu C\epsilon_0 \gamma_0 \Delta}{2\omega^2 
r^2 (r^2+a^2)^{3/2}}e^{-i\omega r_*}.
\label{explicitS}
\end{equation}
Here, $C=\left[\frac{8Z_{l}^{a\omega}}{\sin^2\theta}\right]_{\theta=0}$
and $Z_l^{a\omega}(\theta)$ is a spin-weighted spheroidal harmonic 
\cite{spheroidal}.
The function $\gamma_0=\gamma_0(r)$ can be found in \cite{nmothers}.
The parameter $\epsilon_0$ is the energy per unit rest mass of the 
infalling particle,
which we take to be a very large quantity ($\epsilon_0 
\rightarrow \infty$), 
since we are interested in highly relativistic particles.
The Sasaki-Nakamura equation (\ref{sn}) is to be solved under 
the ``only outgoing
radiation at infinity'' boundary condition, meaning
\begin{equation}
X(\omega,r)=X^{\rm out}e^{i\omega r_*} \,, r_* \rightarrow \infty.
\label{xout}
\end{equation}
Once $X^{\rm out}$ is known, Teukolsky's radial function $R$ can be found,
when $r_* \rightarrow \infty$ (the region of interest here) as
\begin{equation}
R=-\frac{4\omega ^2 X^{\rm out}}{\lambda(\lambda+2)-12i\omega -
12a^2\omega^2}r^3 e^{i\omega r_*} 
=R^{\rm out} r^3 e^{i\omega r_*}.
\label{relationRX}
\end{equation}
Following Nakamura and Sasaki \cite{nmothers} we define the multipolar 
structure so as to have
\begin{equation}
\Delta E=\frac{\pi}{4}\int_{0}^{\infty}\sum_{lm}|h^{lm}(\omega)|^2+
|h^{lm}(-\omega)|^2.
\label{power}
\end{equation}
Once $X$ is known, we can get the Teukolsky wavefunction near infinity 
from (\ref{relationRX}), and the energy radiated away from 
(\ref{power}).

We would now like to find $X(\omega,r)$ from the Sasaki-Nakamura 
differential equation
(\ref{sn}). This is accomplished by a Green's function technique,
constructed so as to satisfy the usual boundary conditions, i.e., 
only
ingoing waves at the horizon ($X \sim e^{-i\omega r_*}
\,,r_*\rightarrow -\infty$) 
and outgoing waves at infinity($X \sim e^{i\omega r_*}
\,,r_*\rightarrow \infty$). 
We get that, near infinity (we are interested in 
knowing the wavefunction in this region),
\begin{equation}
X^{\rm out}=-\frac{\mu\epsilon_0 c_0 C }{4i\omega^3 B}\int 
\frac{e^{-i\omega r_*} X^{H}}{r^2(r^2+a^2)^{1/2}} dr.
\label{solution}
\end{equation}
Here $c_0 \equiv \gamma_0(r=\infty)$, and
$X^{H}$ is an homogeneous solution of (\ref{sn}) which
asymptotically behaves as
\begin{eqnarray}
X^{H} \sim A(\omega)e^{i\omega r_*}+B(\omega)e^{-i\omega r_*}\,,r_* 
\rightarrow \infty \\
X^{H} \sim e^{-i\omega r_*} \,,r_* \rightarrow -\infty \,. 
\label{behavior1}
\end{eqnarray}
We now discuss the numerical results. 

\section{Numerical Results and Conclusions}
The main result of this paper is shown in Fig. 1, the energy spectra
as a function of the angular quantum number $l$, for an ``extreme'' 
black hole with $a=0.999M$. The results for other rotation parameters
$a$ are not shown, both because there are no qualitatives changes and mostly
because they follow from the discussion below.
The most interesting and important features of these results are:

(i) Fig. 1 clearly shows that the zero frequency limit (ZFL) of the
energy spectra, $\frac{dE}{d\omega}_{\omega\rightarrow 0}$ is
non-vanishing.  We know \cite{quem} that the existence of a ZFL is
closely linked to a non-zero velocity at infinity, so this comes as no
surprise and was already observed \cite{lemosvitor} in the collision
of a Schwarzchild black hole with an highly relativistic particle.
The peculiar property of the ZFL arises when we look more closely into
the numerical results: the numerical data shows that the ZFL is
exactly (up to the numerical error) equal to the ZFL for the infall of
an highly relativistic particle into a Schwarzschild black
hole. This means that the ZFL seems to be independent of the spin of
the colliding objects, and given by Smarr's \cite{smarr2} expression.

(ii) The spectra is almost flat up to a critical ($l$-dependent) frequency, a
fact also evident from Fig. 1. Again, this was also true for a
non-rotating black hole \cite{lemosvitor,smarr2}. The spectra is flat
up to $\omega \sim \omega_{QN}$, where $\omega_{QN}$ is the lowest
quasinormal frequency for the Kerr black hole (for work on quasinormal
modes on the Kerr geometry see \cite{QNKerr}, for example).
For $\omega > \omega_{QN}$, the spectra decays exponentially, according
to the empirical law $\frac{dE}{d\omega}\sim e^{-bl}$, ($\omega > \omega_{QN}$), 
first discovered by Davis et al \cite{davis}, and further discussed by 
Sasaki and Nakamura \cite{nmothers}.

\vskip 5mm

\centerline{\epsffile{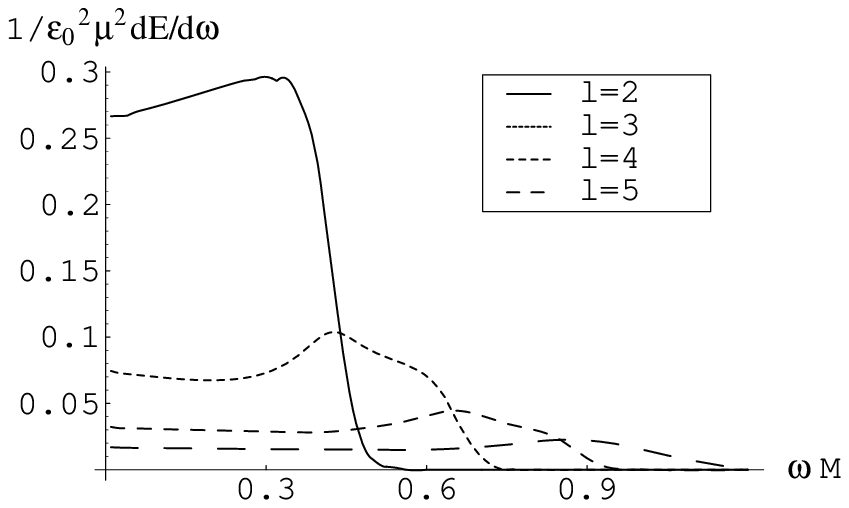}} 
{\noindent {\small Figure
1. The energy spectra for a point particle moving 
at nearly the speed of
light and colliding, along the symmetry axis, with an 
extreme (a=0.999M) Kerr black hole.
Notice that the spectra is almost flat, the ZFL is non-vanishing 
and that the
quadrupole carries less than half of the total radiated energy.}}
\vskip 5mm

(iii) The total energy radiated in each multipole goes as $\Delta E_l
\sim \frac{1}{l^2}$, which (needless to say) is exactly what happens
for Schwarzschild case \cite{lemosvitor}. This power-law dependence
seems to be {\it universal} for highly relativistic collisions, and
was first observed for the non-rotating case \cite{lemosvitor}.  We
note that for collisions beginning at rest, the behaviour is
strikingly different, for the energy goes as $\Delta E_l({\rm rest\,
at\,} \infty) \sim e^{-bl}$ \cite{davis, nmothers}.

(iv) This power-law dependence implies that the radiation is not
quadrupole ($l=2$) in nature, as it was for infall starting from
rest. In fact, less than $50 \%$ of the total energy is carried in the
quadrupole part.

These characteristics plus the discussion in \cite{lemosvitor} allow
one to infer that the total energy radiated will depend mainly on the
behaviour of the quasinormal frequencies for the Kerr geometry.  Now,
for $m=0$, we know \cite{QNKerr} that the quasinormal frequencies for
this geometry are almost the same (but slightly larger) as for the
Schwarzschild geometry, so we expect the total energy to be similar in
both cases (slightly larger in the Kerr case).  In fact, for this
axisymmetric collision we find, on summing over all $l$'s that
\begin{equation}
\Delta E =0.31 \frac{\mu^2\epsilon_0^2}{M}\,\quad\quad a=0.999M\,,
\label{totalenergy}
\end{equation} 
with a $5\%$ error.
We recall that for a Schwarzschild black hole \cite{lemosvitor}
$\Delta E =0.26 \frac{\mu^2\epsilon_0^2}{M}$ so the infall along the
symmetry axis of a Kerr hole does not enhance very much the total
radiated energy (in comparison with the non-rotating case). This was
also observed \cite{nmothers} for collisions along the symmetry axis,
but starting from rest.  As we lower the rotation parameter, the total
energy decreases, and approaches the non-rotating value given above.
What can we say about the collision at nearly the speed of light
between a Schwarzschild and a Kerr black hole, along its symmetry
axis?  Supposing (with all due precautions mentioned in the
Introduction) that (\ref{totalenergy}) holds for $\mu \epsilon_0
\rightarrow M$, we should have an efficiency of $15,5 \%$ for that
process.  This seems a reasonable result, but only a full numerical 
scheme  can tell how
accurate it is. For the moment, one can only say that it is a ``good''
result: it is higher than for the collision of two Schwarzschild black
holes, but still within the upper limit imposed by 
the Area Theorem  $\Delta E \leq 0.66  \, M$ and an eficiency less 
than $33 \%$ \cite{hawking}.
These numbers arise noting that the situation is axisymmetric, the
radiation carries no angular momentum, and thus one has the same 
$a=0.999M$ for the final black hole.

\section*{Acknowledgements}
This work was partially funded by Funda\c c\~ao para a
Ci\^encia e Tecnologia (FCT) through project PESO/PRO/2000/4014. V.C.
also acknowledges finantial support from FCT through PRAXIS XXI
programme.  J. P. S. L. thanks Observat\'orio Nacional do Rio de
Janeiro for hospitality.

%

\end{document}